# Stacking order dynamic in the quasi-two-dimensional dichalcogenide 1*T*-TaS$_2$ probed with MeV ultrafast electron diffraction

L. Le Guyader[1,2*], T. Chase[1,3], A. Reid[1], R.K. Li[1], D. Svetin[4], X. Shen[1], T. Vecchione[1], X.J. Wang[1], D. Mihailovic[4], H.A. Dürr[1•]

*email: llg@slac.stanford.edu
•email: hdurr@slac.stanford.edu

[1] SLAC National Accelerator Laboratory, 2575 Sand Hill Road, Menlo Park, California 94025, USA
[2] European XFEL GmbH, Holzkoppel 4, 22869 Schenefeld, Germany
[3] Department of Applied Physics, Stanford University, Stanford, California 94305, USA
[4] Jozef Stefan Institute and CENN Nanocenter, Jamova 39, SI-1000, Ljubljana, Slovenia

ABSTRACT

Transitions between different charge density wave (CDW) states in quasi-two-dimensional materials may be accompanied also by changes in the inter-layer stacking of the CDW. Using MeV ultrafast electron diffraction, the out-of-plane stacking order dynamics in the quasi-two-dimensional dichalcogenide 1*T*-TaS$_2$ is investigated for the first time. From the intensity of the CDW satellites aligned around the commensurate $l$ = 1/6 characteristic stacking order, it is found out that this phase disappears with a 0.5 ps time constant. Simultaneously, in the same experiment, the emergence of the incommensurate phase, with a slightly slower 2.0 ps time constant, is determined from the intensity of the CDW satellites aligned around the incommensurate $l$ = 1/3 characteristic stacking order. These results might be of relevance in understanding the metallic character of the laser-induced metastable "hidden" state recently discovered in this compound.

INTRODUCTION

Recent developments in the control of charge density waves (CDW), i.e. a combined periodic modulation of the electron density and a periodic lattice distortion, with either electrical current or femtosecond (fs) laser pulses could open the door for novel electronic devices. [Yoshida2015, Cho2016, Ma2016, Liu2016, Vaskivskyi2016, Stojchevska2014] In particular, the 1$T$ polytype of $TaS_2$ displays a rich phase diagram, with the metallic normal (N) phase without CDW existing above T = 543 K, the incommensurate (I) phase with CDW order down to T = 354 K, the nearly commensurate (NC) phase down to T = 183 K and the insulating commensurate (C) phase below.[Ishiguro1991] Upon warming from the commensurate phase, a stripped discommensurate (T) phase is formed at 220 K and remains up to 280 K. In addition to these phases accessible via cycling the sample temperature, a novel metallic metastable "hidden" (H) phase was shown to form upon excitation by a single fs laser pulse. [Vaskivskyi2016, Stojchevska2014] While 1$T$-$TaS_2$ is essentially thought of as 2-dimensional system, with a weak van der Waals interaction between adjacent layers, it is speculated that the insulating and conducting phase properties are driven by the layer stacking order in the 3$^{rd}$ dimension. [Ritschel2015] In the recent years, laser induced dynamics between different phases has been extensively studied with ultrafast electron diffraction (UED) focusing on transitions from the commensurate to nearly commensurate and from the nearly commensurate to incommensurate phases in 1$T$-$TaS_2$ [Eichberger2010, Han2015, Haupt2016] and in 1$T$-$TaSe_2$.[Sun2015] However, all these studies were performed at normal incidence and therefore the stacking order dynamics has remained essentially unexplored.

Here we investigate how and on which time scale the stacking order evolves upon femtosecond laser excitation. Using the 3.3 MeV electron diffraction setup at SLAC, [Weathersby2015] we probe the ultrafast dynamics of the commensurate to incommensurate phase transition upon excitation by a 50 fs, 800 nm wavelength laser pulse. Taking advantage of the nearly flat Ewald sphere for 3.3 MeV electrons, we measured simultaneously the different dynamic behavior of the set of CDW satellite reflections at various wave vectors along the $l$-Bragg rod.

EXPERIMENTAL METHODS

Time-resolved electron diffraction measurements were conducted at the UED setup at SLAC, whose detailed description is given elsewhere.[Weathersby2015] Essentially, 3.3 MeV electron bunches are sent at a 180 Hz repetition rate through a thin 1$T$-TaS$_2$ sample. The diffracted electron beam is detected at a 3 m travel distance after the sample on a phosphor screen, which is imaged with an Andor camera recording the diffraction pattern, as depicted in Fig. 1. The 800 nm wavelength pump laser employed to excite the sample is focused down to a spot size of 1 mm FWHM (Full Width at Half Maximum). At the same time, the electron probe spot size is kept at least 3 times smaller at 300 µm FWHM. The sample temperature can be varied between 35 K to 360 K. The sample can be rotated from its normal by angles up to $\theta$ = +/- 10° without blocking the electron beam by the sample mount, allowing us to investigate the out-of-plane ordering dynamics. Electron diffraction pattern can be recorded at variable pump-probe delay with acquisition times of a few seconds.

The free-standing sample investigated here was prepared by first exfoliating flakes from a TaS$_2$ single crystal sample in a Gel-Pack® box. These flakes were then transferred on a Poly(methyl methacrylate) (PMMA) film on glass. Atomic Force Microscopy (AFM) was used to determine the thickness of these flakes, and the one used in this study was found to be 60 nm thick. The PMMA with TaS$_2$ flakes was then dissolved in acetone and the floating flakes were caught on a Transmission Electron Microscope (TEM) copper grid and finally washed in Isopropyl Alcohol (IPA).

RESULTS AND DISCUSSION

An electron diffraction pattern from our 1$T$-TaS$_2$ sample is shown in Fig. 2a for a sample tilt $\theta$ = 0° from normal incidence and a temperature T = 150 K in the commensurate phase. The first Brillouin zone around the **100** and **200** Bragg peaks are shown as hexagons surrounding them. The 6 first order satellites $q_1$ and -$q_1$, $q_2$ and -$q_2$, $q_3$ and -$q_3$ of the triple CDW in 1$T$-TaS$_2$ are visible around each central Bragg peak and correspond to the star-of-David formation where 13 Ta atoms cluster together, as shown in Fig. 1a. Upon increasing the temperature to 300 K in the nearly commensurate phase, the intensity of all the first order satellites vanishes, as shown in Fig. 2b. To understand this behavior, one should first consider the changes occurring in the basal plane. In the commensurate phase, the CDW satellites form a $\varphi$ = 13.9° angle in the basal plane with the *hk0* reciprocal lattice. This angle changes abruptly at the commensurate to nearly commensurate phase transition by about -2°, after which it slowly reduces with temperature in the nearly commensurate phase until the nearly commensurate to incommensurate phase transition is reached where it changes abruptly to $\varphi$ = 0°.

[Ishiguro1991] However, this rotation of the CDW satellites in the basal plane cannot explain the vanishing of the satellites intensity observed at normal incidence. To understand this, it is necessary to consider the change in out-of-plane stacking order occurring between the commensurate phase and the nearly commensurate phase by tilting the sample. An electron diffraction pattern recorded for a sample tilt of θ = 5° from normal incidence is shown in Fig. 2d for the nearly commensurate phase at T = 300 K. Here, three bright first order satellites, namely $-q_1$, $-q_2$ and $-q_3$ appear around the **200** Bragg peak, which for θ = 5° is near $l$ = 1/3. At the same time, the first order satellite remain very weak around **100** Bragg peak, which for θ = 5° is near $l$ = 1/6. Upon cooling to 150 K in the commensurate phase, this changes as the 3 first order satellites $-q_1$, $-q_2$ and $-q_3$ are now visible around **100** while the ones around **200** become weaker. This behavior suggests a change of stacking order from a $l$ = 1/3 to a $l$ = 1/6.

To determine further the equilibrium out-of-plane stacking order in our sample, the tilt was varied and the averaged intensity of the three positive ($q_1$, $q_2$, $q_3$) and three negative ($-q_1$, $-q_2$, $-q_3$) CDW satellites around **200** are shown in Fig. 3a as dashed orange and blue lines, respectively, as a function of the Miller index $l$ for the nearly commensurate phase at T = 300 K. Here we clearly see a peak around $l$ = 1/3 for the negative CDW satellites and the symmetric behavior with a peak at $l$ = 2/3 = -1/3 for the positive CDW satellites. Following Nakanishi & al. (1984) naming convention, this is in accordance with the <$\sigma_{h2}$> stacking found for the nearly commensurate phase which is a 3 times repeat of the same in-plane translation, depicted in Fig. 1a with the green open circle, connecting the center of a star-of-David formation and the center of the 2-5-6 green triangle. [Nakanishi1984] This arrangement correspond in the reciprocal space to satellite intensity narrowly located around $l$ = 1/3. Upon cooling to the commensurate state as shown in Fig. 3b for T = 150 K, the peaks significantly broaden and shift to $l$ = 1/6 for the negative satellites and to $l$ = 5/6 = -1/6 for the positive satellites with an additional smaller peak at $l$ = 1/3 and $l$ = 2/3 = -1/3 respectively. This is compatible with a change from the ordered <$\sigma_{h2}$> stacking to the partly disordered ($\tau_c$, $\sigma_c$) staking, which is a succession of an in-phase stacking $\tau_c$ followed by a random selection between 3 different translated stacking $\sigma_c$, as depicted in Fig. 1a. [Tanda1984, Nakanishi1984] This demonstrates that for this particular tilt angle θ = 5°, we are both sensitive to the commensurate stacking order with $l$ = 1/6 around the **100** Bragg peak and to the nearly commensurate stacking order with $l$ = 1/3 around **200** Bragg peak, simultaneously.

Next we will address how this change in out-of-plane stacking order dynamically occurs when the system changes from the commensurate to the nearly commensurate phase.

To investigate how this staking order dynamically changes, time-resolved pump-probe measurements were performed near the commensurate to nearly commensurate phase transition at a sample base temperature of 140 K and with a laser pump fluence of $\mathcal{F}$ = 3.0 mJ/cm$^2$. The diffracted intensity distribution around the **100** Bragg peak as function of the azimuth angle χ, as defined in Fig. 2d, at negative delay t = -1 ps and positive delay t = 10 ps, are shown in Fig. 3a. It can be seen here that all the 6 satellites intensity decrease upon excitation. For the $q_3$ (-$q_3$) satellites, the time dependent intensity around the commensurate position $q_{3C}$ (-$q_{3C}$) with φ = 13.9º and incommensurate position $q_{3I}$ (-$q_{3I}$) with φ = 0º are shown in Fig. 3b as open and closed symbols respectively. The intensity at the commensurate position decreases for both $q_3$ and -$q_3$ with a 0.5 ps characteristic time constant and remain constant afterwards up to 12 ps. At the same time no dynamics is observed at the incommensurate position $q_{3I}$ and -$q_{3I}$. From this information alone, one would be tempted to conclude that the laser pulse is weakening the CDW, resulting in a decrease of the diffracted intensity without inducing any particular phase transition.

However, the picture changes drastically when looking at the **200** Bragg peak, for which the satellites profiles before and after the laser pulse excitation are shown in Fig. 3c. It is evident there that the -$q_3$ satellite is both more intense and rotated towards the incommensurate position after laser excitation. This increase of intensity, as opposed to the decrease observed around the **100** Bragg peak, is unambiguously indicative of phase transition, here a change in stacking order from the commensurate phase at $l$ = 1/6 to either the nearly commensurate or incommensurate phase at $l$ = 1/3, the position around which this **200** Bragg peak is aligned. The time dependence of -$q_{3I}$ incommensurate satellite, shown in Fig. 3d, is characterized by a strong increase with a characteristic time constant of 2.0 ps as extracted from a fit with two exponential time constants. At the same time, the -$q_{3C}$ commensurate satellite shows a moderate increase which could be due to the increase seen at the -$q_{3I}$ incommensurate satellite and the limited q-resolution in the recorded UED patterns. The latter broadens the satellites to a peak width that is of the order of the commensurate satellite rotation φ = 13.9º and originates from the electron bunch emittance, sample flatness and crystal quality. The $q_{3C}$ commensurate satellite, which is aligned on the $l$ = 1/3 side peak of the partly disordered ($\tau_c$, $\sigma_c$) staking of the commensurate phase, as shown in Fig. 3b, shows a significant decrease and is therefore another indication that the commensurate phase is disappearing. The $q_{3l}$ incommensurate satellite shows a moderate intensity increase with a faster time constant than for the -$q_{3l}$ incommensurate satellite which seems to be related with the overall increase in diffuse scattering of the excited sample seen as an increased background in Fig. 4c.

CONCLUSION

In summary, using MeV electron diffraction, the stacking order dynamic in 1$T$-TaS$_2$ was for the first time investigated. Thanks to the flat Ewald sphere displayed by MeV electrons, several CDW satellites could be investigated simultaneously, characterizing both the initial commensurate phase and the excited incommensurate phase. We evidenced that the commensurate phase disappears with a characteristic time constant of 0.5 ps while the incommensurate phase emerges with a slower 2.0 ps time constant. It is crucial to realize that all these data are recorded simultaneously and not in essentially different experiments as it would be the case with hard X-rays diffraction for example.[Zhu2015] Moreover, with an improved q-resolution, it would become possible to discern the commensurate and the nearly commensurate satellite intensity as well as the "hidden" phase, opening the path to the study of their in-plane and out-of-plane ordering dynamic and metallic character.

ACKNOWLEDGMENTS

The authors would like to thank SLAC management for their continued support. The technical support by SLAC Accelerator Directorate, Technology Innovation Directorate, LCLS Laser Science & Technology division and Test Facilities Department is gratefully acknowledged. This work is supported by the Department of Energy, Office of Science, Basic Energy Sciences, Materials Sciences and Engineering Division, under Contract No. DE-AC02-76SF00515. Use of the ultrafast Electron diffraction facility at SLAC is supported by the U.S. Department of Energy Contract No. DE-AC02-76SF00515, DOE Office of Basic Energy Sciences Scientific User Facilities Division Accelerator and Detector R&D program, the SLAC UED/UEM Initiative Program Development Fund. L.L.G. thanks the Volkswagen-Stiftung for financial support through the Peter-Paul-Ewald Fellowship.

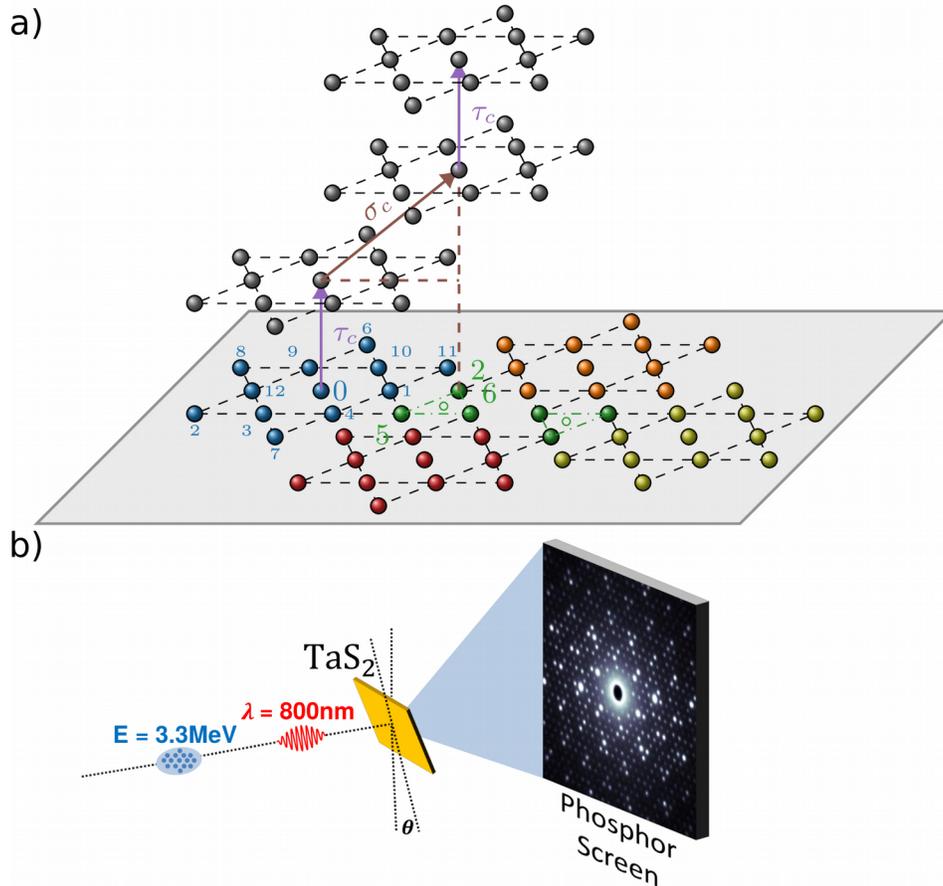

Fig. 1 *a) Direct space view of the star of David formation in the basal plane, where the 13 Ta atoms are numbered in sequence. The out-of-plane layer stacking for the commensurate phase is a succession of a $\tau_c$ translation, positioning the star of David on top of each others in position 0, followed by a random selection between 3 possible $\sigma_c$ translation positioning the star of David on top of position 2, 5 or 6. In the nearly commensurate or incommensurate phase, the $<\sigma_{h2}>$ or $<\sigma_I>$ stacking, respectively, is a three times repeat of a translation linking the center of a star of David and the center of the 2-5-6 green triangle. After three such translation, the layer is back in the 0 position. b) Schematic of the time-resolved electron diffraction experiment, where the sample is tilted by an angle θ from normal incidence to probe the out-of-plane layer stacking changes between the commensurate and nearly commensurate or incommensurate phase. The 800 nm wavelength 50 fs laser pulse excite the TaS2 sample. After a delay t, the 3.3 MeV 100 fs electron bunch diffracts from the sample on a phosphor screen.*

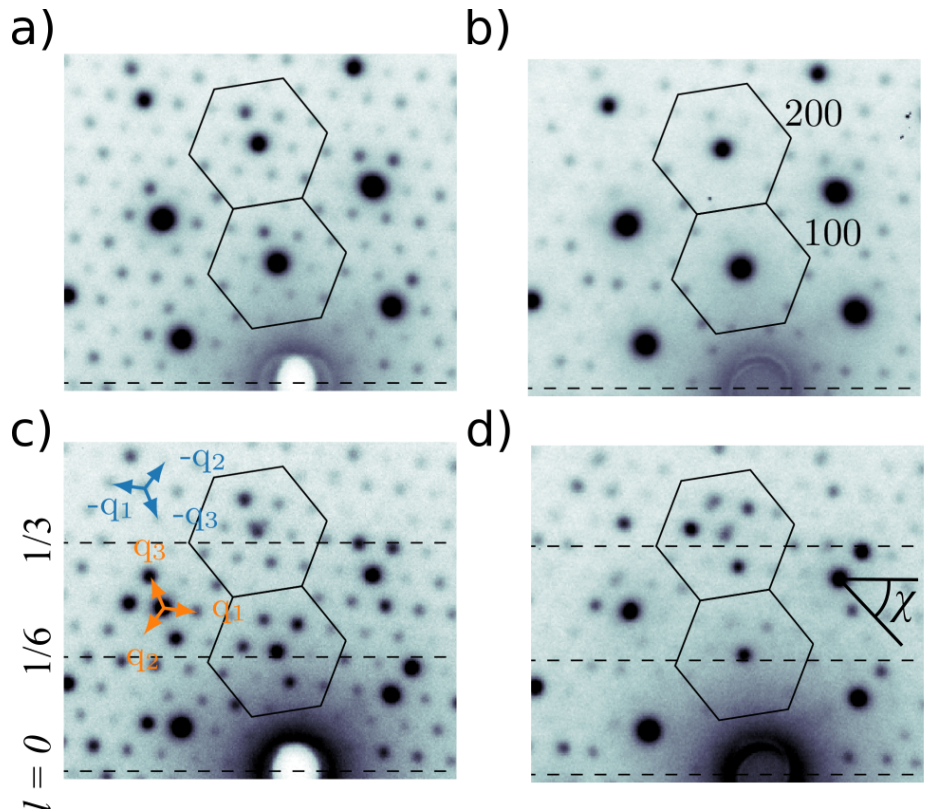

*Fig. 2: part of the electron diffraction pattern recorded at normal incidence a) in the commensurate phase at T = 150 K and b) in the nearly commensurate phase at T = 300 K, as well as for the case of a tilt angle of θ = 5° from normal incidence at c) T =150 K and d) T = 300 K. The first Brillouin zones around the 100 and 200 Bragg peaks are shown as the hexagons. The 3 positive first order satellites $q_1$, $q_2$, $q_3$ are shown as orange arrows while the 3 negative satellites of the CDW $-q_1$, $-q_2$ and $-q_3$ are shown as blue arrows. For the tilted case, dashed lines at l = 0, 1/6 and 1/3 are shown. In d) the azimuth angle χ is shown. In all cases the data are shown on a logarithm scale.*

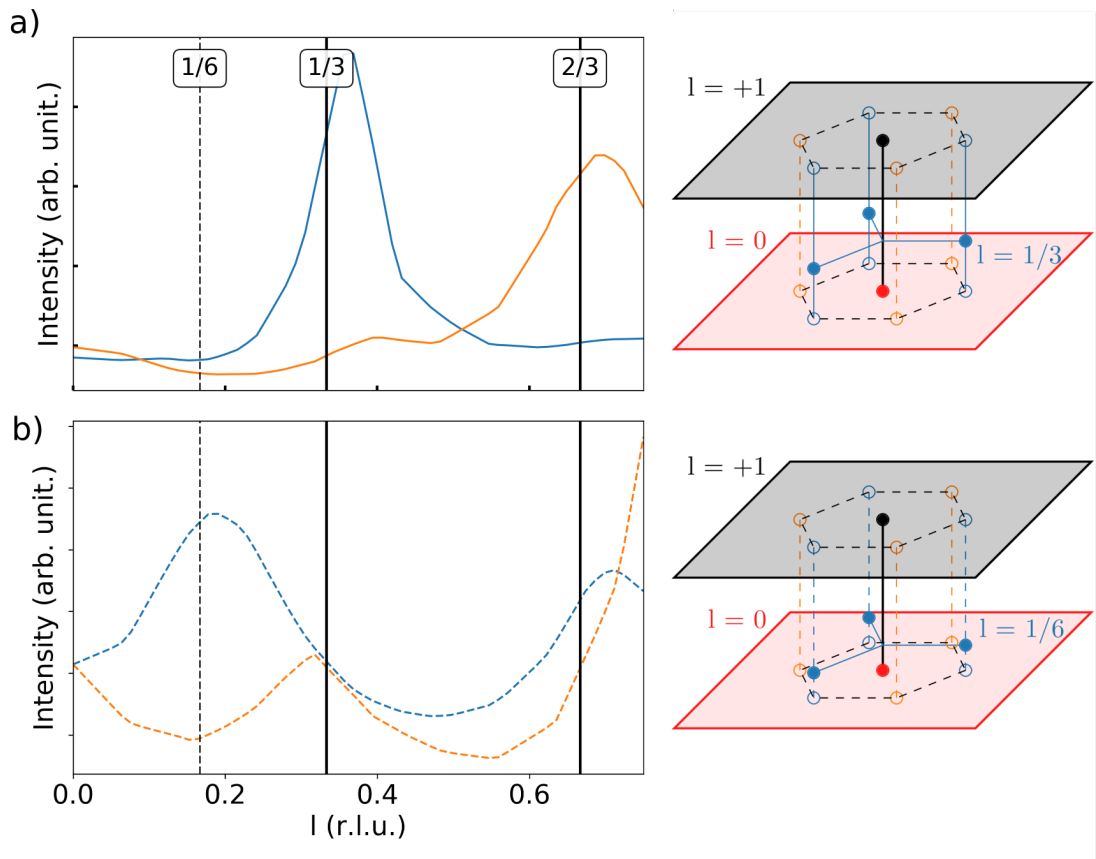

*Fig. 3: Averaged intensity of the 3 positives CDW satellites in orange and 3 negatives satellites in blue as function of the Miller index l for a) the nearly commensurate phase at T = 300 K and for b) the commensurate phase at T = 150 K, together with a schematic in each case of the CDW in reciprocal space with the filled blue satellites at either l = 1/3 or l = 1/6 for the nearly commensurate or commensurate phase respectively. The open orange and blue satellites circles are the projection on the l = 0 and 1 plane which are seen at normal incidence.*

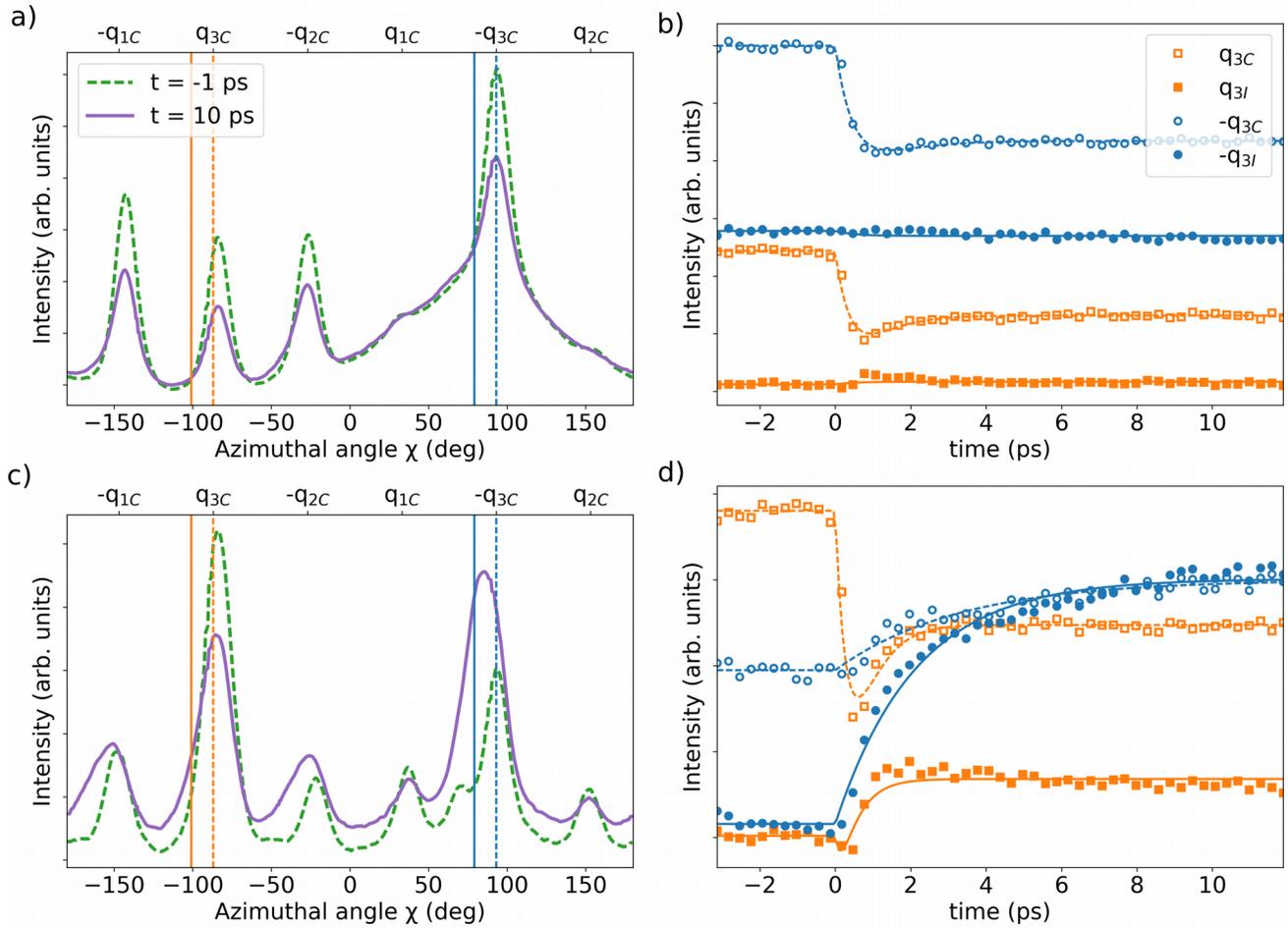

Fig. 4: *a) Intensity profile of the 6 first order CDW satellites around the **100** Bragg peak for a negative delay at t= -1 ps in dashed green line and a positive delay at t = 10 ps in continuous purple line. The vertical dashed blue (orange) line indicates the position in the commensurate phase of the -q3 (q3) satellite -$q_{3C}$ ($q_{3C}$) while the continuous blue (orange) line indicates the position for the incommensurate phase of -$q_3$ (q3) satellite -$q_{3I}$ ($q_{3I}$). b) Time evolution after laser excitation of the CDW satellites intensity at the commensurate (open symbols, dashed lines) and incommensurate position (filled symbols, continuous lines) for the $q_3$ (orange) and -$q_3$ (blue) satellites where the lines corresponds to a fit of the data points with a two exponential time constants. c) and d) are for the **200** Bragg peak. In all figures the sample temperature is T = 140 K and the laser fluence is $\mathcal{F}$ = 3.0 mJ/cm$^2$.*